\documentclass[prd,twocolumn,a4paper,superscriptaddress,floatfix]{revtex4}
\usepackage{graphicx}
\usepackage{bbm}
\usepackage[all]{xy}
\usepackage{amsmath}
\usepackage{amssymb}
\usepackage{epstopdf}

\def\be{\begin{equation}}
\def\ee{\end{equation}}
\newcommand{\bq}{\begin{eqnarray}}
\newcommand{\eq}{\end{eqnarray}}
\newcommand{\bes}{\begin{subequations}}
\newcommand{\ees}{\end{subequations}}

\def\ben{\begin{eqnarray}}
\def\een{\end{eqnarray}}
\def\ba{\begin{array}}
\def\ea{\end{array}}

\begin{document}
\newcommand{\half}{{\textstyle\frac{1}{2}}}
\allowdisplaybreaks[3]
\def\a{\alpha}
\def\b{\beta}
\def\g{\gamma}\def\G{\Gamma}
\def\d{\delta}\def\D{\Delta}
\def\ep{\epsilon}
\def\et{\eta}
\def\z{\zeta}
\def\t{\theta}\def\T{\Theta}
\def\l{\lambda}\def\L{\Lambda}
\def\m{\mu}
\def\f{\phi}\def\F{\Phi}
\def\n{\nu}
\def\p{\psi}\def\P{\Psi}
\def\r{\rho}
\def\s{\sigma}\def\S{\Sigma}
\def\ta{\tau}
\def\x{\chi}
\def\o{\omega}\def\O{\Omega}
\def\k{\kappa}
\def\pa {\partial}
\def\ov{\over}
\def\br{\\}
\def\ud{\underline}

\newcommand\lsim{\mathrel{\rlap{\lower4pt\hbox{\hskip1pt$\sim$}}
    \raise1pt\hbox{$<$}}}
\newcommand\gsim{\mathrel{\rlap{\lower4pt\hbox{\hskip1pt$\sim$}}
    \raise1pt\hbox{$>$}}}
\newcommand\esim{\mathrel{\rlap{\raise2pt\hbox{\hskip0pt$\sim$}}
    \lower1pt\hbox{$-$}}}
\newcommand{\dpar}[2]{\frac{\partial #1}{\partial #2}}
\newcommand{\sdp}[2]{\frac{\partial ^2 #1}{\partial #2 ^2}}
\newcommand{\dtot}[2]{\frac{d #1}{d #2}}
\newcommand{\sdt}[2]{\frac{d ^2 #1}{d #2 ^2}}    


\title{Effective dark energy equation of state in interacting dark energy models}

\author{P.P. Avelino}
\email[Electronic address: ]{ppavelin@fc.up.pt}
\affiliation{Centro de Astrof\'{\i}sica da Universidade do Porto, Rua das Estrelas, 4150-762 Porto, Portugal}
\affiliation{Departamento de F\'{\i}sica e Astronomia da Faculdade de Ci\^encias
da Universidade do Porto, Rua do Campo Alegre 687, 4169-007 Porto, Portugal}

\author{H.M.R. da Silva}
\email[Electronic address: ]{hilberto.silva@gmail.com}
\affiliation{Departamento de F\'{\i}sica e Astronomia da Faculdade de Ci\^encias
da Universidade do Porto, Rua do Campo Alegre 687, 4169-007 Porto, Portugal}

\begin{abstract}

In models where dark matter and dark energy interact non-minimally, the total amount of matter in a fixed comoving volume may vary from the time of recombination to the present time due to energy transfer between the two components. This implies that, in interacting dark energy models, the fractional matter density estimated using the cosmic microwave background assuming no interaction between dark matter and dark energy will in general be shifted with respect to its true value. This may result in an incorrect determination of the equation of state of dark energy if the interaction between dark matter and dark energy is not properly accounted for, even if the evolution of the Hubble parameter as a function of redshift is known with arbitrary precision. In this paper we find an exact expression, as well as a simple analytical approximation, for the evolution of the effective equation of state of dark energy, assuming that the energy transfer rate between dark matter and dark energy is described by a simple two-parameter model. We also provide analytical examples where non-phantom interacting dark energy models mimic the background evolution and primary cosmic microwave background anisotropies of phantom dark energy models.

\end{abstract} 
\maketitle

\section{Introduction}

There is now very strong evidence that our Universe is undergoing a phase of accelerated expansion \cite{Percival:2009xn,Amanullah:2010vv,Komatsu:2010fb}. In the context of general relativity, the only plausible explanation for the present acceleration of the Universe relies on it being dominated by an exotic dark energy (DE) form, violating the strong energy condition \cite{Copeland:2006wr,Frieman:2008sn,Caldwell:2009ix,Li:2011sd}. Observations also indicate that most of the matter in the Universe is non-baryonic and dark. Still, the fundamental nature of both dark matter (DM) and dark DE remains a mystery. It is therefore interesting to consider the possibility of a non-minimal interaction in the dark sector and to investigate the corresponding cosmological implications \cite{Wetterich:1994bg,Amendola:1999er,Zimdahl:2001ar,Farrar:2003uw}. The coupling between DM and DE may affect both the background evolution of the Universe as well as the linear growth of cosmological perturbations \cite{Gumjudpai:2005ry,Pettorino:2008ez,CalderaCabral:2008bx,CalderaCabral:2009ja,Baldi:2010pq}. Interacting dark energy (IDE) may also play an important role on small non-linear scales, potentially affecting the dynamics of galaxies and clusters galaxies \cite{Manera:2005ct,Mainini:2006zj,Bertolami:2007zm,Abdalla:2007rd,Koyama:2009gd,Abdalla:2009mt,Baldi:2010td,Baldi:2011wa,Bertolami:2011yp,Lee:2011tq,Marulli:2011jk,Baldi:2011wy,Avelino:2011zx} (see also \cite{Maccio:2003yk,Baldi:2008ay,Baldi:2010ks,Baldi:2011qi} for a discussion of N-body simulations with IDE). In some models (unified DE) the interaction between DM and DE may be strong enough for the dark sector as a whole to be effectively described by a single fluid \cite{Beca:2005gc,Avelino:2008zz}. IDE models have also been considered as a possible solution to the coincidence problem \cite{TocchiniValentini:2001ty,Cai:2004dk,Berger:2006db,Hu:2006ar,Sadjadi:2006qp} (see however \cite{Olivares:2007rt,Barreira:2011qi}).

The impact of a non-minimal coupling between DM and DE on the Cosmic Microwave Background (CMB) anisotropies, Baryonic Acoustic Oscilations and the apparent magnitude of type Ia supernovae has been investigated by several authors \cite{Amendola:2002bs,Amendola:2003eq,Wang:2006qw,Amendola:2006dg,Guo:2007zk,He:2008tn,Feng:2008fx,Gavela:2009cy,Micheletti:2009pk,Xia:2009zzb,Valiviita:2009nu,Wei:2010uh,Honorez:2010rr,He:2010im,Xu:2011ts,Cao:2011cg,Clemson:2011an} (see also \cite{Martinelli:2010rt,Amendola:2011ie,Beynon:2011hw} for future prospects). Such coupling can lead to a fractional matter abundance significantly different from the one obtained using the information contained in the CMB temperature power spectrum assuming no interaction between DM and DE, which may result in a biased determination of the evolution of the equation of state (EoS) parameter of DE. In this paper we perform an analytical study of a two-parameter model for the energy transfer rate between DM and DE determining, in particular, the impact of ignoring the interaction between DM and DE in the reconstruction of the evolution of the DE EoS parameter. 

Throughout this paper we shall use units with $c=8\pi G/3=H_0=1$, where $c$ is the speed of light in vacuum, $G$ is the gravitational constant, $H$ is the Hubble parameter and the subscript `0' refers to the present time.

\section{Background IDE model}

In a homogeneous and isotropic Friedmann-Robertson-Walker universe the evolution of the matter and DE densities, ($\rho_m$ and $\rho_w$, respectively) in an IDE model is given by
\bq
{\dot \rho}_m&+&3H{\rho_m}=Q \label{intm}\,,\\
{\dot \rho}_w&+&3H(1+w){\rho_w}=-Q \label{intw}\,,
\eq
where a dot represents and derivative with respect to cosmic time $t$, $H={\dot a}/a$ is the Hubble parameter, $a$ is the scale factor, $w=p_w/\rho_w$, $p_w$ is the pressure associated with the DE component. The interaction term will be parametrized by
\be
Q(z)=\alpha H (1+z)^{-\beta} \rho_w \label{qterm}\,,
\ee
where $\beta$ is a constant, $1+z=1/a$, $z$ is the cosmological redshift, $a_0$ is assumed to be equal to unity throughout the paper (the subscript `0' refers to the present time or, equivalently, $z=0$). If $\alpha$ is taken to be a function of $a$ then this term is completely general. However, in the present paper we shall assume that $\alpha$ is a constant in order to be able to find analytical solutions. For simplicity, we shall only consider the toy model case with $w={\rm constant}$ (see \cite{Avelino:2009ze,Avelino:2011ey} for a discussion of such models). 

Eq. (\ref{intw}) implies that the DE density is
\be
\rho_w=\Omega_{w0} f(z) (1+z)^{3(1+w)}\,,
\ee
where
\be
\frac{df}{dz}=\alpha f(z) (1+z)^{-\beta-1}\label{dfdz}\,,
\ee
$\Omega_{w}=\rho_w/\rho_c$, $\rho_c=H^2$ is the critical density and $f_0\equiv f(0)=1$. The exact solution to Eq. (\ref{dfdz}) is given by 
\be
f(z)=\exp\left(-\frac{\alpha}{\beta}\left[ \left(1+z\right)^{-\beta}-1\right]\right)\,, \label{fexact}
\ee
if $\beta \neq 0$, or 
\be
f(z)=(1+z)^{\alpha}\,,
\ee
if $\beta=0$.

\begin{figure}
\includegraphics[width=8.5cm]{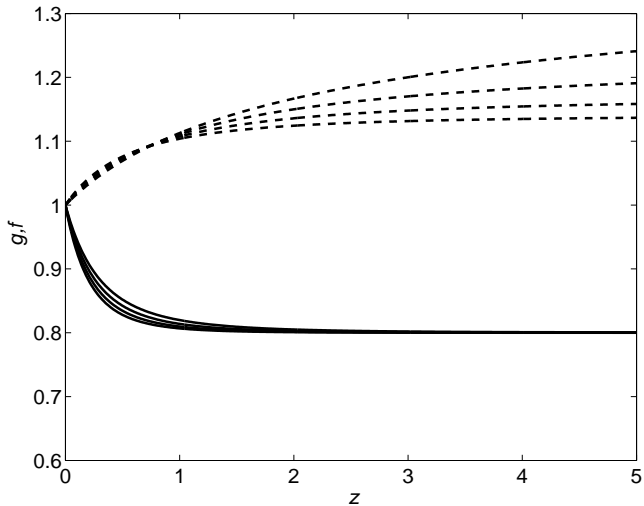}\caption{Evolution of $g$ (solid line) and $f$ (dashed line) with redshift $z$ for $\beta=0.5,1,2,4$, $w=-0.99$ and $\Omega_{m0} g_\infty = 0.27$. The functions $f(z)$ and $g(z)$ were computed using Eqs. (\ref{fexact}) and (\ref{gexact}), respectively, and the values of $\alpha$ were chosen such that $g_\infty=0.8$ for all models.
\label{fig1}}
\end{figure}

From now on we shall only consider the case with $\beta > 0$ which is the one that reproduces the standard evolution of the matter and DE energy densities early on (that is $\rho_m \propto (1+z)^{3}$ and $\rho_w \propto (1+z)^{3(1+w)}$ at large redshift). If
\be
\left|\frac{f_\infty-f_0}{f_0}\right|=\left|f_\infty-1\right| \ll 1\label{approx}\,,
\ee
or equivalently if $|\alpha/\beta| \ll 1$, then the solution in Eq. (\ref{fexact}) is given approximately by
\be
f(z)=1-\frac{\alpha}{\beta}\left[ \left(1+z\right)^{-\beta}-1\right]\label{approx1}\,,
\ee
with $f_\infty \equiv f(\infty)=1+\alpha/\beta$.

Eq. (\ref{intm}) implies that the matter density can now be written as
\be
\rho_m=\Omega_{m0}g(z)(1+z)^3 \,,
\ee
where $g(z)$ satisfies the equation
\be
\frac{dg}{dz}=-\alpha \frac{\Omega_{w0}}{\Omega_{m0}} f(z) (1+z)^{\gamma}\label{dgdz}\,,
\ee
with $\gamma=3w-\beta-1$ (note that if $-1 \le w \le 0$ and $\beta >0$ then $-4 < \gamma< -1$). The exact solution to Eq. (\ref{dgdz}) is given by
\bq
&\ &g(z)=1-\frac{\Omega_{w0}}{\Omega_{m0}} 
\left(\frac{\alpha}{\beta}\right)^{(\gamma+\beta+1)/\beta} \exp \left(\frac{\alpha}{\beta}\right) \times \nonumber \\
&\times& \left( \Gamma \left(-\frac{\gamma+1}{\beta},\frac{\alpha}{ \beta(1+z)^{\beta}}\right)- \Gamma \left(-\frac{\gamma+1}{\beta},\frac{\alpha}{\beta}\right)\right)\ \label{gexact}
\eq
where 
\be
\Gamma(x,y)=\int_y^\infty t^{x-1} e^{-t} dt\,,
\ee
is the incomplete gamma function.

\begin{figure}
\includegraphics[width=8.5cm]{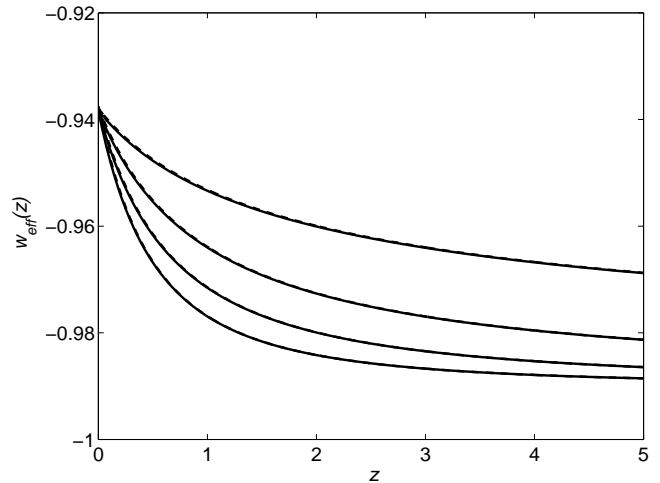}\caption{Evolution of the effective DE EoS parameter $w_{eff}$ with redshift $z$ computed using the exact solution obtained from Eqs. (\ref{fexact}), (\ref{gexact}) and (\ref{waexact}) for $\beta=0.5,1,2,4$, $w=-0.99$ and $\Omega_{m0} g_\infty = 0.27$ (solid lines) as well as the corresponding evolution obtained using the approximation given in Eq. (\ref{waapprox}) for the same values of $\beta$, $w$ and $\Omega_{m0}$ (dashed lines). The values of $\alpha$ were chosen such that $g_\infty=0.8$ for all models.
\label{fig2}}
\end{figure}

Fig. 1 shows the evolution of $g$ (solid line) and $f$ (dashed line) with redshift $z$ for $\beta=0.5,1,2,4$, assuming that $w=-0.99$, $\Omega_{m0} g_\infty = 0.27$ (with $g_{\infty} \equiv g(\infty)$) and $\Omega_{w0}=1-\Omega_{m0}$. The functions $f(z)$ and $g(z)$ were computed using Eqs. (\ref{fexact}) and (\ref{gexact}), respectively, and the values of $\alpha$ were chosen such that $g_\infty=0.8$ for all models. The convergence towards the asymptotic values at high redshift of both $g$ and $f$ is an increasing function of $\beta$.  Fig. 1 shows that $g$ converges faster towards its asymptotic value at high redshift than $f$ due to the fact that, in these models, the DE density at high redshift represents only a very small fraction of the total density.

\begin{figure}
\includegraphics[width=8.5cm]{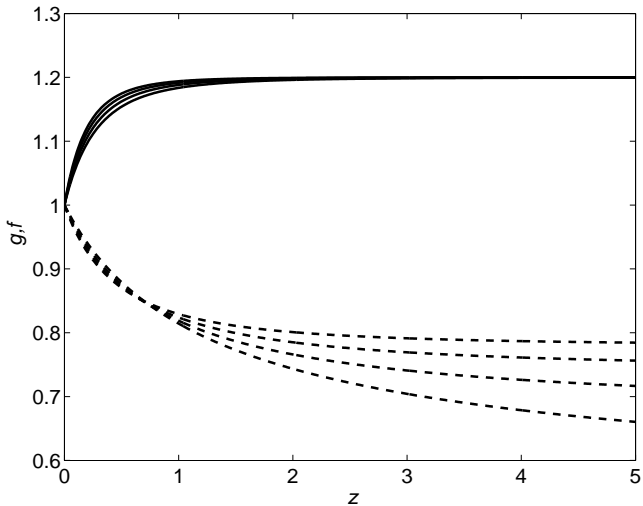}\caption{Same as Fig. 1 but with $g_{\infty}=1.2$ and $w=-0.95$.
\label{fig3}}
\end{figure}

It will turn out to be useful to find a first order (in $\alpha/\beta$) approximate solution to Eq. (\ref{dgdz}). If the first order solution for $f(z)$ given in Eq. (\ref{approx1}) is substituted in Eq. (\ref{dgdz}) one finds
\bq
g(z)&=&1- \frac{\alpha}{\gamma+1} \frac{\Omega_{w0}}{\Omega_{m0}} \left(\left(1+\frac{\alpha}{\beta}\right)\left[(1+z)^{\gamma+1}-1\right]\right. - \nonumber\\
&-&\left.\frac{\alpha}{\beta}C\left[(1+z)^{\gamma-\beta+1}-1\right] \right)\label{gzfirst}\,,
\eq
where 
\be
C=\frac{\gamma+1}{\gamma-\beta+1}\,,
\ee
with $0<C<1$, so that
\bq
g_\infty&=&1+\frac{\alpha}{\gamma+1}   \frac{\Omega_{w0}}{\Omega_{m0}} \left(1+\frac{\alpha}{\beta}\frac{\beta}{\beta-\gamma-1}\right)=\nonumber\\
&=&1-\frac{\alpha}{\beta-3w}   \frac{\Omega_{w0}}{\Omega_{m0}} \left(1+\frac{\alpha}{2\beta-3w}\right)\,.
\eq
Up to first order in $\alpha/\beta$ one has
\be
g_\infty = 1- \frac{\alpha}{\beta-3w} \frac{\Omega_{w0}}{\Omega_{m0}}\,,
\ee
and
\bq
\Delta g(z) &\equiv& g(z)-g_\infty = (1-g_\infty) \times (1+z)^{\gamma+1}=\nonumber\\
&=&\frac{\alpha}{\beta-3w} \frac{\Omega_{w0}}{\Omega_{m0}}(1+z)^{-\beta+3w}\,.\label{Deltag}
\eq
Note that in Eq. (\ref{gzfirst}) the $(1+z)^{\gamma+1}\ge(1+z)^{\gamma-\beta+1}$ for any $z \ge 0$ (assuming that $\beta > 0$).

\begin{figure}
\includegraphics[width=8.5cm]{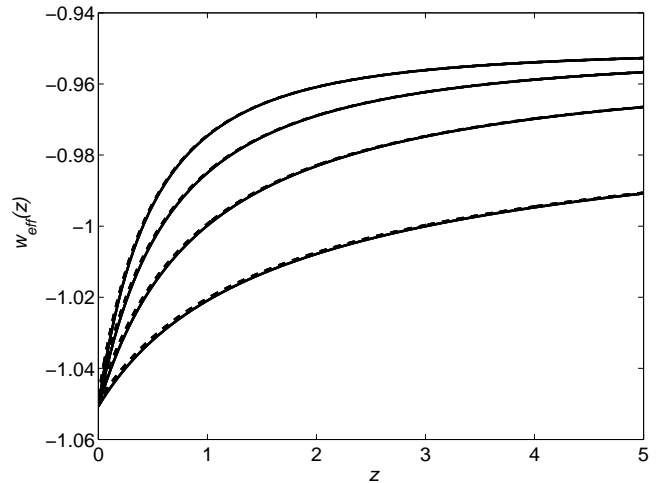}\caption{Same as Fig. 2 but with $g_{\infty}=1.2$ and $w=-0.95$.
\label{fig3}}
\end{figure}

\section{Effective DE EoS parameter}

The background evolution in a flat Friedmann-Robertson-Walker universe may be determined from the equations (in a flat universe $\Omega_{w0}=1-\Omega_{m0}$)
\bq
H^2 &=& \rho=\Omega_{m0} g(z) (1+z)^3+\nonumber \\
&+& (1-\Omega_{m0}) f(z) (1+z)^{3(w+1)} \,, \\
-\frac23 {\dot H}&=&\rho+p= \Omega_{m0} g(z) (1+z)^3+\nonumber\\
&+&(1+w)(1-\Omega_{m0}) f(z) (1+z)^{3(w+1)} \,.
\eq
The pressure of the matter component is assumed to be negligible and consequently the pressure of the DE component may be computed as 
\bq
p&=&p_w(z)=-\frac23 {\dot H}-H^2=H\left(\frac23 \frac{dH}{dz}(1+z)-H\right)=\nonumber\\
&=&  w f(z)(1-\Omega_{m0}) (1+z)^{3(w+1)}\,.
\eq
Primary cosmic microwave background temperature anisotropies constrain the fractional matter ratio at $z_{rec} \sim 10^3$. In this paper we shall consider that $\beta$ is large enough so that the energy transfer between DM and DE is completely negligible at recombination or, equivalently, that $g(z_{rec})$ is very close to $g_\infty$. This way the physics of recombination is (almost) unaffected by the non-minimal interaction between DM and DE, apart from the fact that in IDE models the matter density at recombination is given by
\be
\rho_m(z_{rec})=g_\infty \Omega_{m0}(1+z_{rec})^3\,,
\ee
rather than $\Omega_{m0}(1+z_{rec})^3$.
Hence, if the non-minimal coupling between DM and DE is not taken into account, the fractional matter density at the present day is estimated as $g_\infty \Omega_{m0}$ rather than $\Omega_{m0}$. This implies that the DE density would also be (wrongly) estimated as
\bq
\rho_{weff}(z)&=&H^2-g_{\infty} \Omega_{m0}(1+z)^{3}=\Omega_{m0} \Delta g(z) (1+z)^3+\nonumber\\
&+& (1-\Omega_{m0}) f(z) (1+z)^{3(w+1)}\,,
\eq
corresponding to an effective DE EoS parameter given by
\bq
w_{eff}&=&\frac{p_w}{\rho_{weff}}=-\frac{2{\dot H}/3+H^2}{H^2-g_{\infty} \Omega_{m0}(1+z)^{3}}=\nonumber\\
&=&\frac{w}{1+\frac{\Omega_{m0}}{1-\Omega_{m0}} \frac{\Delta g(z)}{f(z)} (1+z)^{-3w}}\,. \label{waexact}
\eq
Using Eqs. (\ref{approx1}) and (\ref{Deltag}) one finds the first order approximation
\be
w_{eff}(z)=w\left(1-\frac{\alpha}{\beta-3w}(1+z)^{-\beta}\right)\,,\label{waapprox}
\ee
with
\bq
w_{eff0} &\equiv& w_{eff}(0)=w\left(1-\frac{\alpha}{\beta-3w}\right)=\nonumber\\
&=&w\left(1+(g_\infty-1)\frac{\Omega_{m0}}{1-\Omega_{m0}}\right)\label{wa0approx}\,.
\eq
Hence, up to first order in $\alpha/\beta$, $w_{eff0}$ is the same for all models with identical values of $w$, $\Omega_{m0}$ and $g_\infty$. Also note that $w_{eff} \to w$ for $z \to \infty$. Interestingly, the first approximation for the evolution of the DE EoS parameter derived in the present paper (Eq.  (\ref{waapprox})) also holds if the parametrization of the DM-DE interaction given in Eq. (\ref{qterm}) is generalized to
\be
Q(z)=\alpha \frac{\Omega_{w0}^{1-\nu_2}}{\Omega_{m0}^{\nu_1}}H (1+z)^{-3(w+1)(\nu_2-1)-3\nu_1-\beta} \rho_m^{\nu_1} \rho_w^{\nu_2} \label{qterm1}\,,
\ee
where $\nu_1$ and $\nu_2$ are constants.

Fig. 2 shows the evolution of the effective value of the DE EoS parameter $w_{eff}$ with redshift $z$ computed using the exact solution obtained from Eqs. (\ref{fexact}), (\ref{gexact}) and (\ref{waexact}) for $\beta=0.5,1,2,4$, $w=-0.99$ and $\Omega_{m0} g_\infty = 0.27$ (solid lines), as well as the corresponding evolution obtained using the approximation given in Eq. (\ref{waapprox}) for the same values of $\beta$, $w$ and $\Omega_{m0}$ (dashed lines). The values of $\alpha$ were chosen such that $g_\infty=0.8$ for all models. Due to Eq. (\ref{wa0approx}) all the models have approximately the same value of $w_{eff0}$ but for larger values of $\beta$ the effective DE EoS parameter $w_{eff}$ converges faster towards its asymptotic value at high redshift ($w$). Fig. 2 shows that the first order approximation given in Eqs. (\ref{waapprox}) is excellent. 

Figs. 3 and 4 are identical to Figs. 1 and 2, respectively, except for the values $w$ and $g_\infty$ ($w=-0.95$ and $g_\infty=1.2$ in the case of Figs. 3 and 4). Similarly to Fig. 1, Fig. 3 shows that the convergence towards the asymptotic values at high redshift is much faster for the function $g$ than for the function $f$ (again the convergence towards  the asymptotic values at high redshift of both $g$ and $f$ is faster for larger values of $\beta$). Fig. 4 shows that, although $w=-0.99 > -1$, the effective EoS parameter becomes smaller than $-1$ at redshifts not too large, a behavior usually associated with phantom DE. This shows that interacting dark energy models may mimic phantom behavior in the absence of phantom dark energy \cite{Das:2005yj}.

\section{Conclusions \label{conc}}

In this paper we investigated IDE models where the energy transfer rate between DM and DE is described, at the background level, by a simple two-parameter model. We used this model to determine the evolution of the effective DE EoS parameter $w_{eff}$ obtained using CMB observations to constrain $\Omega_{m0}$, assuming no interaction between DM and DE and a perfect knowledge of the evolution of the Hubble parameter $H$ with redshift $z$. We found a simple first order approximation to $w_{eff}(z)$, showing that it may be significantly different from the true DE EoS parameter (which was assumed to be equal to a constant $w$) specially at low redshifts ($w_{eff} \to w$ for $z \to \infty$). We also provided some specific analytical examples where non-phantom (IDE) models mimic the background evolution and small scale cosmic microwave background anisotropies of phantom dark energy models.

Although cosmological data sensitive to the growth of density perturbations at low redshifts may be able to distinguish between models with the same $w_{eff}(z)$, it turns out the background energy transfer between DM and DE does not uniquely determine the energy-momentum transfer at a perturbative level. Hence, additional assumptions about the energy-momentum transfer between DM and DE at a perturbative level \cite{Honorez:2010rr,Clemson:2011an} need to be made in order to further constraint IDE models.

\begin{acknowledgments}

This work is partially supported by FCT-Portugal through project CERN/FP/116358/2010.

\end{acknowledgments}


\bibliography{I}

\end{document}